\documentclass[12pt]{iopart}

\def\lsim{\mathrel{\hbox{\rlap{\hbox{\lower4pt\hbox{$\sim$}}}\hbox{$<$}}}}
\def\gsim{\mathrel{\hbox{\rlap{\hbox{\lower4pt\hbox{$\sim$}}}\hbox{$>$}}}}

\newcommand{\citealt}{\cite}
\newcommand{\citep}{\cite}
\newcommand{\citet}{\cite}

\usepackage{citesort}
\usepackage{epsfig}
\usepackage{esint}

\usepackage{xcolor}

\begin{document}

\title[Quasars as sources of the stochastic gravitational wave background]{Can quasars, triggered by mergers, account for NANOGrav's stochastic gravitational wave background?}

\author{\'Agnes Kis-T\'oth$^1$, Zolt\'an Haiman$^{2,3}$ and Zsolt Frei$^1$}

\address{$^1$E\"otv\"os University, Institute of Physics, P\'azm\'any P. s. 1/A, Budapest 1117, Hungary\\$^2$ Department of Astronomy, Columbia University, New York, NY 10027, USA\\$^3$ Department of Physics, Columbia University, New York, NY 10027 USA}
\ead{agnes.kis-toth@ttk.elte.hu, zoltan@astro.columbia.edu and frei@ttk.elte.hu}
\vspace{10pt}
\begin{indented}
\item[]April 2024
\end{indented}

\begin{abstract}
The stochastic gravitational wave background (GWB) recently discovered by several pulsar timing array (PTA) experiments is consistent with arising from a population of coalescing super-massive black hole binaries (SMBHBs). The amplitude of the background is somewhat higher than expected in most previous population models or from the local mass density of SMBHs.  SMBHBs are expected to be produced in galaxy mergers, which are also thought to trigger bright quasar activity.   Under the assumptions that (i) a fraction $f_{\rm bin}\sim 1$ of all quasars are associated with SMBHB mergers, (ii)  the typical quasar lifetime is $t_{\rm Q} \sim 10^8~{\rm yr}$, and (iii) adopting Eddington ratios $f_{\rm Edd}\sim 0.3$ for the luminosity of bright quasars, we compute the GWB associated directly with the empirically measured quasar luminosity function (QLF).  This approach bypasses the need to model the cosmological evolution of SMBH or galaxy mergers from simulations or semi-analytical models. 
We find a GWB amplitude approximately matching the value measured by NANOGrav. 
Our results are consistent with most quasars being associated with SMBH binaries and being the sources of the GWB, and imply a joint constraint on $t_{\rm Q}$, $f_{\rm Edd}$ and the typical mass ratio $q\equiv M_2/M_1$.
The GWB in this case would be dominated by relatively distant $\sim10^9{\rm M_\odot}$ SMBHs at $z\approx 2-3$, at the peak of quasar activity.   Similarly to other population models, our results remain in tension with the local SMBH mass density.
\end{abstract}

\section{Introduction}
\label{sec:introduction}

Several different pulsar timing array (PTA) campaigns have recently reported the discovery of a stochastic gravitational background (GWB) at nano-Hz frequencies, including 
the North American Nanohertz Observatory for Gravitational Waves (NANOGrav; \citealt{NANOGrav15-GWB}),
the joint European PTA and Indian PTA~(EPTA and InPTA; \citealt{EPTA+InPTA-GWB}), 
the Australian Parkes PTA (PPTA; \citealt{Parkes-GWB}), and 
the Chinese PTA~(CPTA; \citealt{ChinesePTA-GWB}).
Although many different physical explanations have been put forward for its origin, the background can be naturally attributed to the cosmological population of coalescing supermassive black hole (SMBH) binaries~\citep{NANOGrav15-SMBHBs}.    The presence of SMBHs in the nuclei of most major galaxies is well established~\citep{KormendyHo2013}, and galaxies are known to be built up by mergers.
Each merger event is expected to deliver the nuclear SMBHs (e.g. \citep{Springel+2005,Robertson+2006}), along with a significant amount of gas (e.g. \citep{BarnesHernquist1992}), to the central regions of the new post-merger galaxy. A natural conclusion is that pairs of SMBHs should frequently form in galactic nuclei over cosmic time-scales, and that this should often take place in gas-rich environments.

With the assistance of the surrounding stars and gas, the pair of SMBHs is widely expected to form a gravitationally bound sub-parsec binary~(e.g.~\citealt{Begelman+1980}; see also a recent review by~\citealt{Bogdanovic+2022}), eventually producing strong GWs and merging.  At the same time, a long-standing suggestion, separate from the existence or nature of SMBH binaries, is that such galaxy mergers trigger nuclear activity~\citep{KauffmannHaehnelt2000}, and may be the main mechanism activating and driving the cosmological evolution of the brightest quasars~\citep{Menci+2014}.   

This last conclusion remains poorly established, both theoretically and observationally, since AGN activity can also be triggered by instabilities inside galaxies and other mechanisms that produce rapid gas inflows towards the nucleus (e.g.~\citep{Sharma+2024} and references therein).  Nevertheless, here we take this connection between quasars and merging SMBHs at face value, and directly compute the $z=0$ GWB from the observed quasar luminosity function (QLF), employing only a minimal set of simple assumptions.   This approach bypasses the need to model the cosmological evolution of SMBH or galaxy mergers from simulations and/or semi-analytical models. Instead, it requires three ingredients: (i) a relation between quasar luminosity and SMBH mass (or "Eddington ratio" $f_{\rm Edd}$), (ii) the fraction of quasars associated with SMBH binaries ($f_{\rm bin}$), and (iii) assuming a quasar lifetime $t_{\rm Q}$ so that the QLF can yield an estimate of the underlying SMBH merger rate.  The purpose of this paper is to assess the values of ($f_{\rm Edd},f_{\rm bin},t_{\rm Q}$) which would produce a GWB consistent with the amplitude recently inferred from PTAs.   The idea of predicting the population of SMBH binaries directly from the QLF through this method was proposed in~\cite{Haiman+2009} and similar approaches have been applied to the GWB in the past~\cite{Sesana+2018,Casey-Clyde+2022,Casey-Clyde+2024}.   

The rest of this paper is organized as follows.
In \S~\ref{sec:methods}, we describe our adopted quasar luminosity function and discuss how it can be used to produce an estimate of the GWB.
In \S~\ref{sec:results}, we show our main results, namely the GWB amplitude as a function of the asumed model parameters.
In \S~\ref{sec:discussion}, we discuss these results, and compare our approach and findings to related recent other studies.
Finally, in \S~\ref{sec:conclusions}, we summarize the main implications and offer our conclusions.

\section{Methods}
\label{sec:methods}

In this section, we first describe the quasar luminosity function (QLF) we adopted, and then discuss how we use it to produce an estimate of the mass function of binary SMBHs and the stochastic GWB.  For pedagogical reasons, we present two different ways to construct the latter estimate - versions of both of these appear in the literature, and here we show that these two ways to compute the GWB are mathematically identical.

\subsection{Binary SMBH number density from the QLF}
\

The number density of luminous quasars as a function of their absolute magnitude and redshift is given by the quasar luminosity function (QLF). The QLF describes the number of luminous quasars in a given (comoving) volume element and magnitude range,

\begin{equation}
    \Phi(M,z) = \frac{d^2N_q(M,z)}{dVdM}.
\end{equation}

There are many different approaches in the literature to determine the shape and evolution of the QLF. Here we adopt the result of \citep{Xin+2021} that is based on the compilation of observations by \citep{Kulkarni+2019}. Taking a sample of more than $80\, 000$ colour-selected quasars from redshift $z = 0$ to $7.5$, the (rest-frame) UV quasar luminosity function was found to be well described by a double power-law at all redshifts:

\begin{equation}
    \Phi(M,z) = \frac{\Phi^*(z)}{10^{(0.4[\alpha(z)+1)(M-M_*(z)]}+ 10^{(0.4[\beta(z) +1)(M-M_*(z)]}}.
\end{equation}
The four parameters of the double power-law, i.e. the overall normalisation $\Phi^*$ given in units ${\rm cMpc^{-3}mag^{-1}}$, the break absolute magnitude $M_{*}$ and the bright– and faint– end slopes, $\alpha$ and $\beta$, all vary with redshift as described in \cite{Xin+2021}.

All quasars are assumed to harbour a supermassive black hole (SMBH). Whether it is a single black hole or a binary, its mass can be related to the quasar's luminosity. Ref.~\cite{Inayoshi+2020} found that assuming a constant (product of the) bolometric correction and the Eddington ratio $f_{\rm Edd}\equiv L/L_{\rm Edd}=1$, the black hole's mass ($M_\bullet$) 
can be estimated from the quasar's absolute rest-frame $1450$\AA\, magnitude ($M_{1450}$) as
\begin{equation}
    \log_{10}(M_\bullet) = \frac{-M_{1450}-3.46}{2.5}.
\end{equation}
A more realistic approach is to assume a distribution of $f_{\rm Edd}$ values. First we examine cases where the quasars are shining at sub-Eddington rates, but $f_{\rm Edd}$ still has a fixed constant value. For different Eddington ratios the quasar's $1450$ \rm{\AA}  magnitude can be given in the form:
\begin{equation}
    M_{1450} = -2.5\log_{10}(M_\bullet\times f_{\rm Edd})-3.46.
\end{equation}
We also investigate cases where $f_{\rm Edd}$ has a probability distribution, adopted from~\cite{Kollmeier+2006}. This study found that the distribution of Eddington ratios can be described as log-normal, with a peak at $f_{\rm Edd} \simeq 1/4$ and a dispersion of $0.3$~dex in the redshift range $z \sim 0.3-4$. 
Based on the mass-magnitude relation and the QLF, the number density $\Psi(M_\bullet,z)$ of SMBHs in the centers of quasars as a function of the black hole's mass and redshift is given by
\begin{equation}
    \Psi(M_\bullet,z)  =\Phi(M_{1450}(M_\bullet),z) \times \frac{dM_{1450}}{dM_\bullet}.
\end{equation}

We make the assumption that all quasars we see on the sky have been actived by a merger event. 
Alternatively, if a fraction $f_{\rm bin}<1$ of quasars are associated with mergers and binary SMBHs, then our predictions below, including the GWB amplitude, scale linearly with  $f_{\rm bin}$.
At any given moment, the number of quasars on the sky should then be given by integrating the activation rate (equal to the merger rate by assumption) for the lifetime for which quasars are detectable. Further assuming that the merger rate is constant over the typical quasar lifetime of $t_{\rm Q}\approx 10^8$yr (which is much shorter than the Hubble time),
we have $\Psi=\dot{n}t_{\rm Q}$. The merger rate density can then be expressed simply as:

\begin{equation} \label{eq:mrd}
    \dot{n}(M_\bullet,z) = \frac{\Psi(M_\bullet,z)}{t_{\rm Q}} =    \frac{\Phi(M_\bullet,z)}{t_{\rm Q}}  \times \frac{dM_{1450}}{dM_\bullet}.
\end{equation}

\begin{figure}[t]
\includegraphics[width=\textwidth]{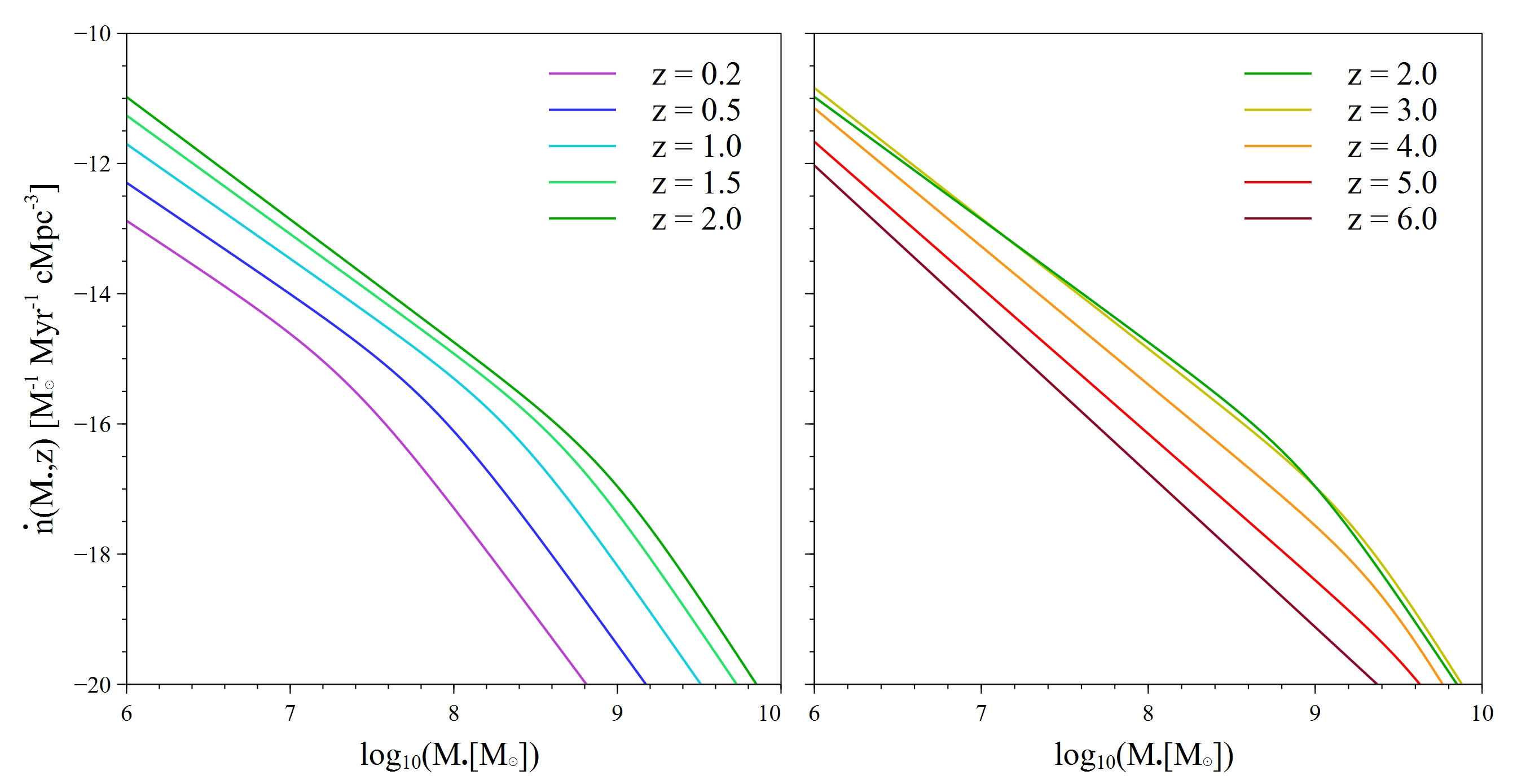}
\caption{The SMBHB merger rate per unit comoving volume as a function of the binary's total mass for nine different redshifts. The merger rate was calculated based on the double  power-law quasar luminosity function (QLF) with the assumption that all quasars have been activated by a merger event, and that quasars are accreting at the Eddington limit. In all cases, the number of merger events are dominated by the least massive binaries with $10^6-10^8 {\rm M_{\odot}}$. The left panel shows that in the $z<2$ range, the merger rate first increases with redshift and then slows down. The right panel shows that in the redshift range $z \sim 2-3$ the merger rate density is almost unchanged, but starts to decrease toward higher redshifts.}
\end{figure}

\vspace{\baselineskip}

We next calculate the stochastic gravitational wave background in two conceptually different ways.  First, we sum the contributions of individual quasars to obtain the total GW "flux" on the sky, or the corresponding {\it surface brightness} $S_{\rm gw}$  of the sky, which can then be converted to a GWB energy density or strain amplitude.  In the second, conceptually somewhat more direct method, we follow \cite{Phinney2001} and integrate the GW luminosity density to directly obtain {\it energy density $\epsilon_{\rm gw}$} in GWs at $z=0$.   These two approaches yield the same result, recalling the relation $\epsilon_{\rm gw}=(4\pi/c) S_{\rm gw}$.

\subsection{Method 1: GWB by summing actively inspiraling individual sources on the sky}\

Our goal is to sum the contributions of individual SMBH binaries associated with luminous quasars on the sky. The contribution of a single binary system to the sky- and polarization-averaged GW strain can be expressed from the GW luminosity of circular binaries

\begin{equation}
    h_{\rm s}^2(f)= \frac{32}{5c^8} \frac{(G \mathcal{M})^{10/3}}{d_{c}^2} (2\pi f_{ p})^{4/3}.
\end{equation}
Here $c$ is the speed of light, $G$ is the gravitational constant, $d_c$ is the comoving distance to the source at redshift $z$, $f_{\rm p}$ is the rest-frame orbital frequency, which is half of the gravitational wave frequency for a circular binary, and $\mathcal{M}$ is the chirp mass.

\begin{equation}
    f_{p} = \frac{f_{r}}{2} =\frac{f(1+z)}{2}, \hspace{1cm}
    \mathcal{M} = \frac{(m_1m_2)^{3/5}}{(m_1 + m_2)^{1/5}}
\end{equation}
The $m_1$ and $m_2$ are the individual masses of the SMBHs, $m_1 + m_2=M_\bullet$ the total black hole mass of the binary, 
and $f$ is the GW frequency observed on Earth.
An individual binary's contribution as a function of its chirp mass, redshift and observed frequency can be calculated as

\begin{equation}
    h_{\rm s}^2(\mathcal{M},z,f)= \frac{32}{5c^8} \frac{(G \mathcal{M})^{10/3}}{d_{c}^2(z)} [\pi (1+z)f]^{4/3}
\end{equation}

We first assume, in the simplest scenario, that all SMBH binaries are inspiraling due to energy loss only through gravitational wave radiation. The frequency evolution during the gravitational wave driven inspiral in the rest frame of a circular binary with given chirp mass is given by

\begin{equation}
    \frac{df_{r}}{dt_{r}} = \frac{96 \pi^{8/3} G^{5/3}}{5 c^5} \mathcal{M}^{5/3}f_{r}^{11/3}.
\end{equation}
Since the total duration of observable quasar activity 
(measured in the quasar's rest frame), once triggered by a merger event, is $t_{\rm Q}$, the probability of catching a binary in a given logarithmic frequency interval can be calculated 
from the ratio of the residence time the binary spends in this frequency band $t_{\rm res}=f/\dot{f}$ to $t_{\rm Q}$,

\begin{equation}
    \frac{dP}{d\ln f_{r}}(\mathcal{M},f_{r}) = 
    \frac{t_{\rm res}}{t_{\rm Q}}=
    \frac{f_{r}}{t_{\rm Q} (df_{r}/dt_{r})}.
\end{equation}
The probability depends on the chirp mass and the rest-frame frequency, so from our point of view on Earth it depends on the redshift as well,

\begin{equation}
    \frac{dP}{d\ln f}(\mathcal{M},z,f) = \frac{dP}{d\ln f_r}(\mathcal{M},(1+z)f).
\end{equation}
The relation between the binary system's chirp mass and the total black hole mass in the system can be given by the mass ratio  $q\equiv m_2/m_1 \leq 1$ as

\begin{equation}
    \mathcal{M}= M_\bullet\left[\frac{q}{(q+1)^2} \right]^{3/5}.
\end{equation}
We first assume a constant $q=1$ for simplicity, and below we investigate cases with different constant values, and also calculate our results for different probability distributions of $q$. In the latter case, the function $\Psi(m,z)$ must be defined in such a way that it gives the number density of SMBHBs depending not only on the total mass but also on the $q$ parameter. In this case $\Psi(m,z)$ must be weighted with the distribution of $q$, and we denote it as $\Psi_{q}(m,q,z)$.

Finally we combine the number density of binary SMBHs associated with quasars, the probability of catching a binary in a given logarithmic frequency interval and the GW "flux" from individual binary systems. In this case the gravitational strain, equivalent to the total GW flux in a logarithmic observed frequency bin $d\ln f$, summing over all sources on the sky at any given moment can be calculated as:

\begin{equation}
    h_{\rm c}^2(f) = \iiint dM_{\bullet}\,dq\,dz \,\, \Psi_q(M_{\bullet},q,z) \frac{dV}{dz} \frac{dP}{d\ln f}(\mathcal{M},z,f) \times h_{\rm s}^2(\mathcal{M},z,f).
\end{equation}
Since $\Psi_q$ gives the number density of binaries per unit volume we need to convert it to per redshift bin. The relation between the volume and the redshift through cosmic history can be expressed as:

\begin{equation}
    \frac{dV}{dz} = \frac{4\pi c}{H_0}\frac{d_c^2}{E(z)},
\end{equation}
where $H_0$ is the Hubble constant and $E(z)$ is the dimensionless Hubble parameter. Substituting the values of the functions, we have the following result for the gravitational wave strain:

\begin{equation}
    h_c^2(f) = \frac{4G^{5/3}}{3\pi^{1/3}c^2 H_0} \frac{1}{f^{4/3}}
    \iiint dM_{\bullet}\,dq\,dz\, \frac{\Psi_q(M_{\bullet},q,z)}{E(z)(1+z)^{4/3}} \times \frac{\mathcal{M}^{5/3}}{t_{Q}} \,.
\label{eq:gwb1}
\end{equation}

\subsection{Method 2: GWB by summing the GW energy injected by all past mergers}

We next use the merger rate density and integrate the energy dumped into the universe through gravitational waves in an (observed) logarithmic frequency bin over time as

\begin{equation}
    \epsilon_{\rm gw}(f) = \iint dM_{\bullet}\,dt \,\, \dot{n}(M_{\bullet},z) \times \frac{dE_{f}(\mathcal{M})}{d\ln f_{r}} \times \frac{1}{1+z}.
\end{equation}
Here $\dot{n}$ denotes the merger rate per comoving volume, $dE_f(\mathcal{M})$ is the energy emitted by a binary with chirp mass $\mathcal{M}$, in its rest frame,  into a logarithmic frequency bin $\ln f_r$, and the additional factor of $(1+z)^{-1}$ accounts for the redshifting of GWs between emission and observation.
Note that $dE_f(\mathcal{M})$ is evaluated at the rest-frame frequency $f_r=(1+z)f$. We can link the merger rate back to the number density of SMBHs in the centers of quasars using Equation \ref{eq:mrd} and then we have:

\begin{equation}
    \epsilon_{\rm gw}(f) = \iint dM_{\bullet}\,dz\, \frac{\Psi(M_{\bullet},z) }{t_{\rm Q}} \frac{dt_r}{dz} \times \frac{dE_f(\mathcal{M})}{df_r}f_r  \times \frac{1}{1+z} 
\end{equation}
Since the emitted GW energy depends on the binary's chirps mass and not just on the total mass, here we may also need to take into account the $q = m_2/m_1 \leq 1$ parameter in the number density function of SMBHBs.

\begin{equation}
    \epsilon_{\rm gw}(f) = \iiint dM_{\bullet}\,dq\,dz\, \frac{\Psi_q(M_{\bullet},q,z) }{t_{\rm Q}} \frac{dt_r}{dz} \times \frac{dE_f(\mathcal{M})}{df_r}f_r  \times \frac{1}{1+z} 
\end{equation}
Here the relation between the rest-frame time evolution and the redshift can be calculated as:

\begin{equation}
    \frac{dt_r}{dz} =\frac{1}{H_0(1+z)E(z)}
\end{equation}
And the GW energy emitted in gravitational waves by an inspiraling circular binary with chirp mass $\mathcal{M}$ in a rest-frame frequency bin $f_r$ is:

\begin{equation}
    \frac{dE_f(\mathcal{M})}{df_r} = \frac{\pi}{3G}\frac{(G \mathcal{M})^{5/3}}{(\pi f_r)^{1/3}}.
\end{equation}
The characteristic GW strain $h_{\rm c}$ in a logarithmic frequency band can be calculated for a given chirp mass population using the relation between the characteristic strain and energy density.

\begin{equation}
    h_{\rm c}^2(f) = \frac{4G}{\pi c^2 f^2} \epsilon_{\rm gw}(f)
\end{equation}
That leads to the following equation:
\begin{equation}
    h_c^2(f) = \frac{4G}{\pi c^2 f^2}
    \iiint dM_{\bullet}\,dq\,dz\, \frac{\Psi_q(M_{\bullet},q,z) }{(1+z)t_{\rm Q}} \frac{dt_r}{dz} \times \frac{dE_f(\mathcal{M})}{df_r}f_r  
\end{equation}
Inserting the values of the displayed functions we get the equation: 
\begin{equation}
    h_c^2(f) = \frac{4G^{5/3}}{3\pi^{1/3}c^2 H_0} \frac{1}{f^{4/3}}
    \iiint dM_{\bullet}\,dq\,dz\, \frac{\Psi_q(M_{\bullet},q,z)}{E(z)(1+z)^{4/3}} \times \frac{\mathcal{M}^{5/3}}{t_{Q}} \,.
    \label{eq:gwb2}
\end{equation}
This is equivalent to Eq.~\ref{eq:gwb1} we obtained in the previous subsection.

\subsection{The effect of SMBHBs' gas-driven inspiral for the GWB}\
\label{subsec:gas-driven}

Environmentally-driven binary evolution can be described in many ways, here we adopt the method introduced in \citep{NANOGrav15-SMBHBs}. 
In this model the evolution of the binary's semi-major axis $a$ is parameterized 
with a double power-law function as
\begin{equation}
    \frac{da}{dt}\bigg|_{\rm{Env}} = H_a 
    \left(\frac{a}{a_{\rm{c}}}\right)^{1-\nu_{\rm{inner}}}
    \left(1 + \frac{a}{a_{\rm{c}}}\right) ^{\nu_{\rm{inner}}-\nu_{\rm{outer}}},
\end{equation}
where $a_c$ is the critical separation, $\nu_{\rm inner}$ and $\nu_{\rm outer}$ controls the hardening rate 
of the binaries in the 'inner' (small separation) and 'outer' (large-separation) regime, and 
$H_a$ is the normalization that depends on the total lifetime of the binary denoted as $\tau$. 
When we include GW emission as well, the hardening rates are added linearly:
\begin{equation}
    \frac{da}{dt} = \frac{da}{dt}\bigg|_{\rm Env} + \frac{da}{dt}\bigg|_{\rm GW}.
\end{equation}
The total lifetime of the binary can be calculated based on the following equation.
\begin{equation}
	\tau = \int_{a_{\rm init}}^{a_{\rm isco}} \left( \frac{da}{dt} \right)^{-1} \, da,
\end{equation}
where $a_{\rm init}$ is the initial binary separation and $a_{\rm isco} \equiv 6\, GM/c^2$ is the innermost stable circular orbit, where we consider the binary to have merged. As in \citep{NANOGrav15-SMBHBs} we assume fixed values for the parameters
\begin{equation}
    \nu_{\rm outer} = +2.5, \hspace{5mm} a_c = 100 \, \textrm{pc}, \hspace{5mm} a_{\rm init} = 1000 \, \textrm{pc} .
\end{equation}
The free parameters in our model are  $\nu_{\rm inner}$ and $\tau$. In this paper, we examined the ranges of 
\begin{equation}
     - 0.1 \,<\, \nu_{\rm inner}  \, < \, -1.0\, , \hspace{5mm}  0.1 \, \textrm{Gyr}\,<\, \tau \,<\, 2 \, \textrm{Gyr}.
\end{equation}

\begin{figure}[t]
\includegraphics[width=\textwidth]{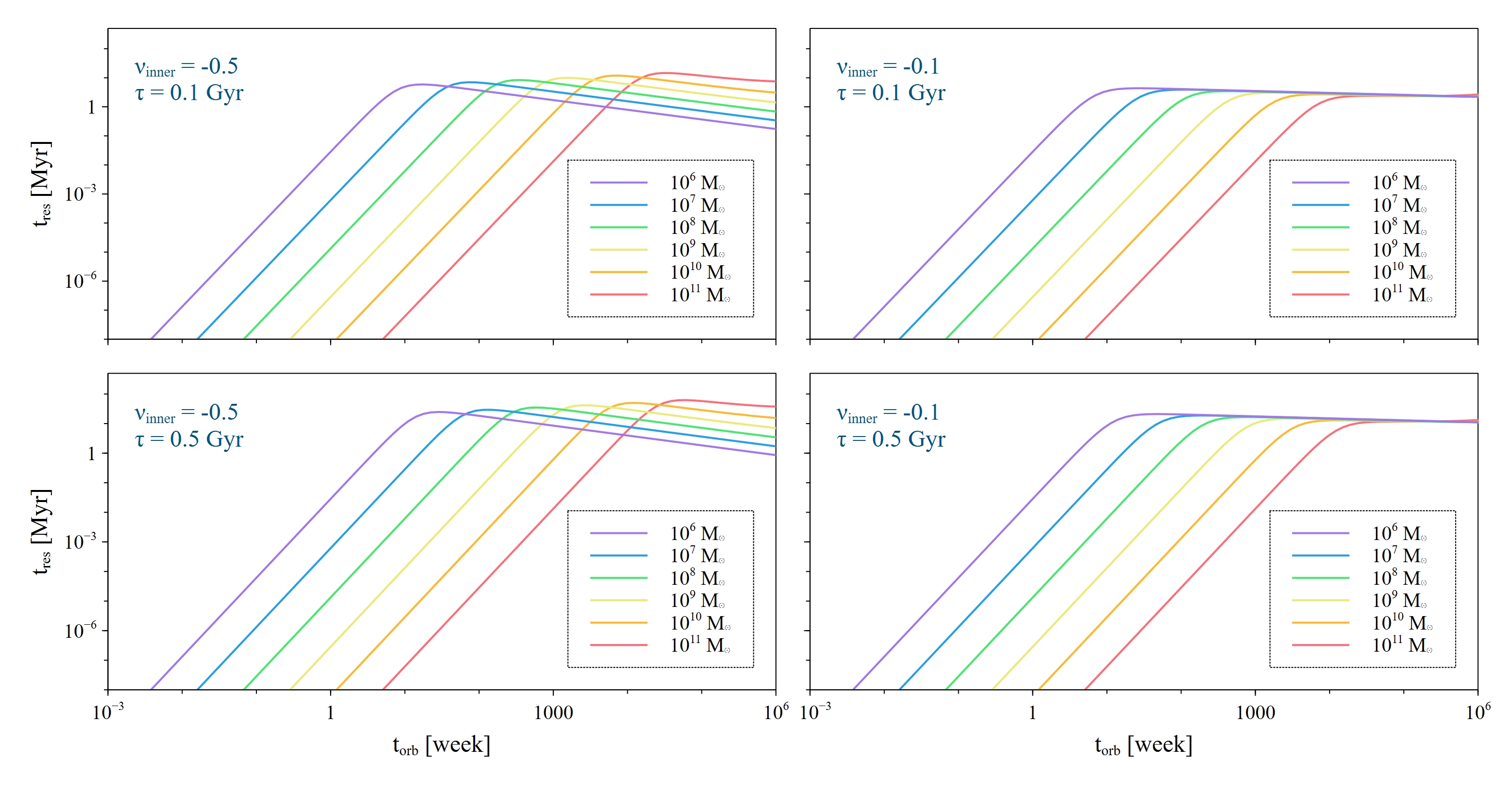}
\caption{The evolution of the SMBHBs' separation for different parameter values of the environmentally-driven inspiral model. All four panels show the residence time ($t_{res} \equiv - a / ( da / dt) $) that the binary spends at orbital separation $a$ as a function of the orbital time ($t_{orb}$) corresponding to that separation. The coloured curves correspond to SMBH binaries with total masses of $M_{\bullet} = 10^{6-11}\,{\rm M_{\odot}}$ as labeled. The different panels show $\nu_{\rm inner} = -0.1 \, , \, -0.5$ for the inner-region separation parameter, and $\tau = 0.1\,,\, 0.5$ Gyr for the binary's total lifetime.} 
\label{fig:inspiral}
\end{figure}

In Figure $\ref{fig:inspiral}$, we show the residence time ($t_{\rm res}$) that a binary spends at a certain orbital separation $a$ corresponding to a given orbital time ($t_{\rm orb}$). In each panel the curves correspond to SMBH binaries with total masses of $M_{\bullet} = 10^{6-11}\,{\rm M_{\odot}}$ for different combinations of the parameters.

\section{Results}
\label{sec:results}

In this section we present our results for the characteristic strain ($h_{\rm c}$) of the gravitational wave background, based on the assumption that all quasars have been activated by a merger event. We define a fiducial model by the chosen values (or distributions) of the parameters $f_{\rm Edd}$, $q$ and $t_{\rm Q}$.
We adopt the results of \citep{Kollmeier+2006} and we use a log-normal distribution for the Eddington ratio, with a peak at $f_{\rm Edd} = 0.25$ and a dispersion of $0.3$~dex. Similarly we apply a log-normal distribution for the mass-ratio $q\equiv m_2/m_1 \leq 1$ following  \citep{Sesana+2018}, with a peak at $q \simeq 1$ and a dispersion of $1.2$~dex. The quasar lifetime in the fiducial model is a constant $t_{\rm Q}\simeq 2.7 \times 10^7$ yr, consistent with expectations from observations~\citep{Martini2004}, and chosen to fit the NANOGrav data (see below).

\subsection{The GW background spectrum}

\begin{figure}[t]
\includegraphics[width=\textwidth]{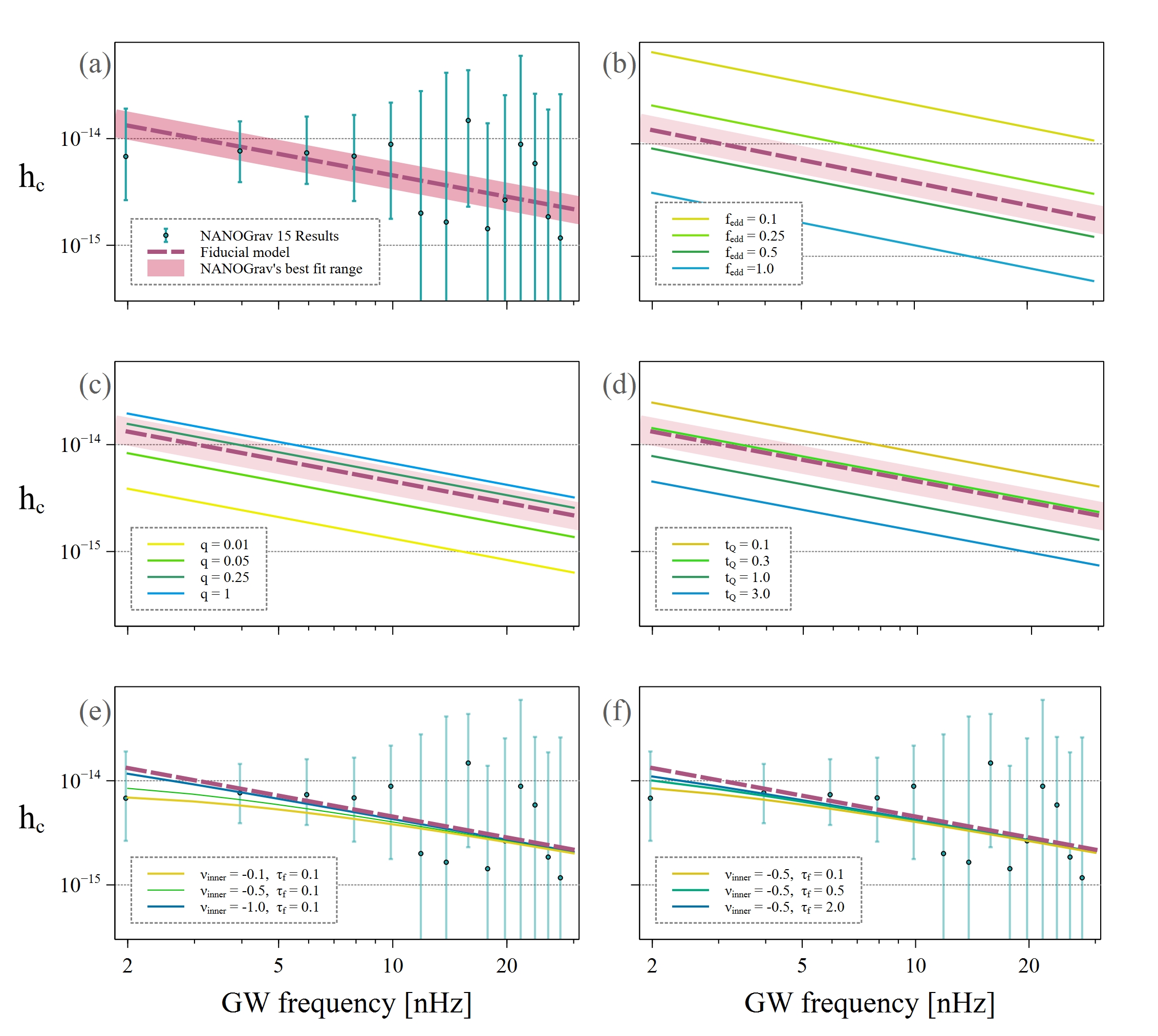}

\caption{Predicted and observed GWB spectra. Panel (a) shows the NANOGrav 15-year results (datapoints with errors bars), and their best-fit power-law spectrum assuming canonical GW-driven SMBH binary inspirals (pink lane, showing the $\pm1\sigma$ range), compared to our fiducial model with $\textrm{log}(f_{\rm Edd}) \sim N(-0.6,0.3)$, $\textrm{log}(q) \sim N(0,1.2)$ and $t_{\rm Q} = 2.7 \times 10^7 \rm \,yr$ (red dashed line).
Panels (b), (c) and (d) show the dependence of the GWB amplitude 
on the Eddington ratio, binary mass ratio, and quasar lifetime, respectively. These illustrate that a lower $f_{\rm Edd}$, higher $q$, and a lower $t_{\rm Q}$ increase the expected GWB. In panels (e) and (f), we take into account the binaries' gas-driven inspiral with different $\nu_{\rm inner}$ (inner slope) and $\tau$ (total binary lifetime), showing that these parameters (defined in \S~\ref{subsec:gas-driven}) impact the spectrum at low frequencies ($\lsim$10 nHz).} 
\label{fig:gwspectra}
\end{figure}

Our calculations for the GWB spectrum based on Eq.~\ref{eq:gwb1} and Eq.~\ref{eq:gwb2}, are shown in Figure~\ref{fig:gwspectra}. In the top left panel, we show a comparison between the NANOGrav data (points with error bars) and our fiducial model (red dashed line). In~\citep{NANOGrav15-GWB} many theoretical models are discussed, including the GWB spectra produced by the SMBHB population. Assuming pure GW-driven orbital evolution and neglecting observational biases (e.g. from stochastic sampling of high-mass binaries), the spectrum can be fit by a power-law:
\begin{equation}
    h_c( f ) = A_{\rm yr} \times \left(\frac{f}{\rm yr^{-1}}\right)^{-\alpha},
\end{equation}
where $A_{\rm yr}$ is the GWB amplitude referenced at a frequency of $1 \, {\rm yr}^{-1}$, and in the idealized case the power index $\alpha = 2/3$. 
The only free parameter is then the amplitude, for which the NANOGrav analysis yields
\begin{equation}
    h_c( f ) = 2.4 \times 10^{-15} \times \left(\frac{f}{\rm yr^{-1}}\right)^{-2/3} = 2.2 \times 10^{-14} \times \left(\frac{f}{\rm nHz}\right)^{-2/3}.
\end{equation}
Our fiducial model reproduces this best-fit power-law GWB. The pink shades around the fiducial model indicate the 90\% confidence range from NANOGrav.
On panel (a) we can see the characteristic strain function generated by the fiducial model, compared to NANOGrav's best fit power-law range and the original data.

Panels (b), (c) and (d) show the effect of changing the values of the parameters in our model. In panel $(b)$ we assume a fixed Eddington ratio for all binaries $f_{\rm Edd} = 0.1, 0.25, 0.5$ and $1.0$, showing that a decrease in $f_{\rm Edd}$  increases the GWB amplitude. A lower Eddington ratio means that the SMBH of a quasar with a magnitude is more massive, producing stronger GWs. In reality, $f_{\rm Edd}$ follows a broad distribution, such as the log-normal distribution in the fiducial model; we only used fixed $f_{\rm Edd}$ to demonstration this dependence.

Panel (c) in Figure \ref{fig:gwspectra} similarly shows the dependence on the mass ratio $q$. Here $q$ was set to different constant values as $q = 0.01, \, 0.05,\, 0.25,\, 1.0 $. It is apparent that a smaller $q$ causes a decrease in the GWB amplitude, as expected, since for a fixed total mass, the chirp mass is lower.

In panel (d), we illustrate the dependence on the lifetime $t_{\rm Q}$. A shorter lifetime means requires a higher intrinsic SMBH abundances to reproduce the observed QLF~\citep{HaimanHui2001,MartiniWeinberg2001}, which consequently increases the GWB amplitude. Similarly to $f_{\rm Edd}$, in reality $q$ and $t_{\rm Q}$ both follow broad distributions.
Nevertheless, the above results show that with a suitable and reasonable combination of these the parameters $f_{\rm edd},\,q$ and $t_{\rm Q}$, the amplitude of the GWB measured by NANOGrav can be reproduced. 

Finally, in the bottom row of Figure~\ref{fig:gwspectra} we demonstrate the impact of environmentally-driven inspirals on the GWB spectrum. We adopt our fiducial model for  $f_{\rm edd},\,q$ and $t_{\rm Q}$ but we modify the residence time to include an environmentally-accelerated early inspiral phase. These results depend on the two additional variables introduced in \S~\ref{subsec:gas-driven}. Panel (e) shows the effect of the inner region inspiral parameter $\nu_{\rm inner}= -0.1,\, -0.5,\, -1.0$, while panel (f) shows the dependence on the total binary lifetime $\tau = 0.1, 0.5, 2.0$. Both of these have a significant impact only at low frequencies ($\approx$2-10 nHz), and all of the cases shown are compatible with the PTA measurements.  Improved measurements and a more detailed investigation in the future may further narrow the constraints on these variables.

\subsection{Parameter degeneracies}

As shown in Figure \ref{fig:gwspectra} the gravitational wave background observed by the NANOGrav can be reproduced by our fiducial model. Given the degeneracies between the model parameters, other combinations of parameter values can be equally allowed. We therefore next investigate our parameter space to find these degenerate combinations. As it is more likely that the parameters have distributions rather than fixed values, we kept the log-normal shapes for the distributions of $f_{\rm edd}$ and $q$. We set the peaks of these distributions to be free parameters, but fixed their dispersions at $0.3$~dex and $1.2$~dex, respectively. For simplicity, we treat $t_{\rm Q}$ as a single free parameter. The three parameters and the corresponding ranges were:

\begin{table}[ht]
    \centering
    \begin{tabular}{c c}
        Peak of the logarithmic Eddington-ratio distribution: & $\, -2.0 < \textrm{log}({f_{\rm Edd}}^*) < 0.0 $\\
        Peak of the logarithmic mass-ratio distribution: & $\, -3.0 < \textrm{log}(q^*) < 0.0 $ \\
        Logarithmic quasar lifetime: & $\, -2.0 \, < \textrm{log}(\, t_{\rm Q} \,/\, 10^8 \, \textrm{yr} \,)< 1.0 \,$
    \end{tabular}
    \label{tab:parameters}
\end{table}

\begin{figure}[t]
\includegraphics[width=\textwidth]{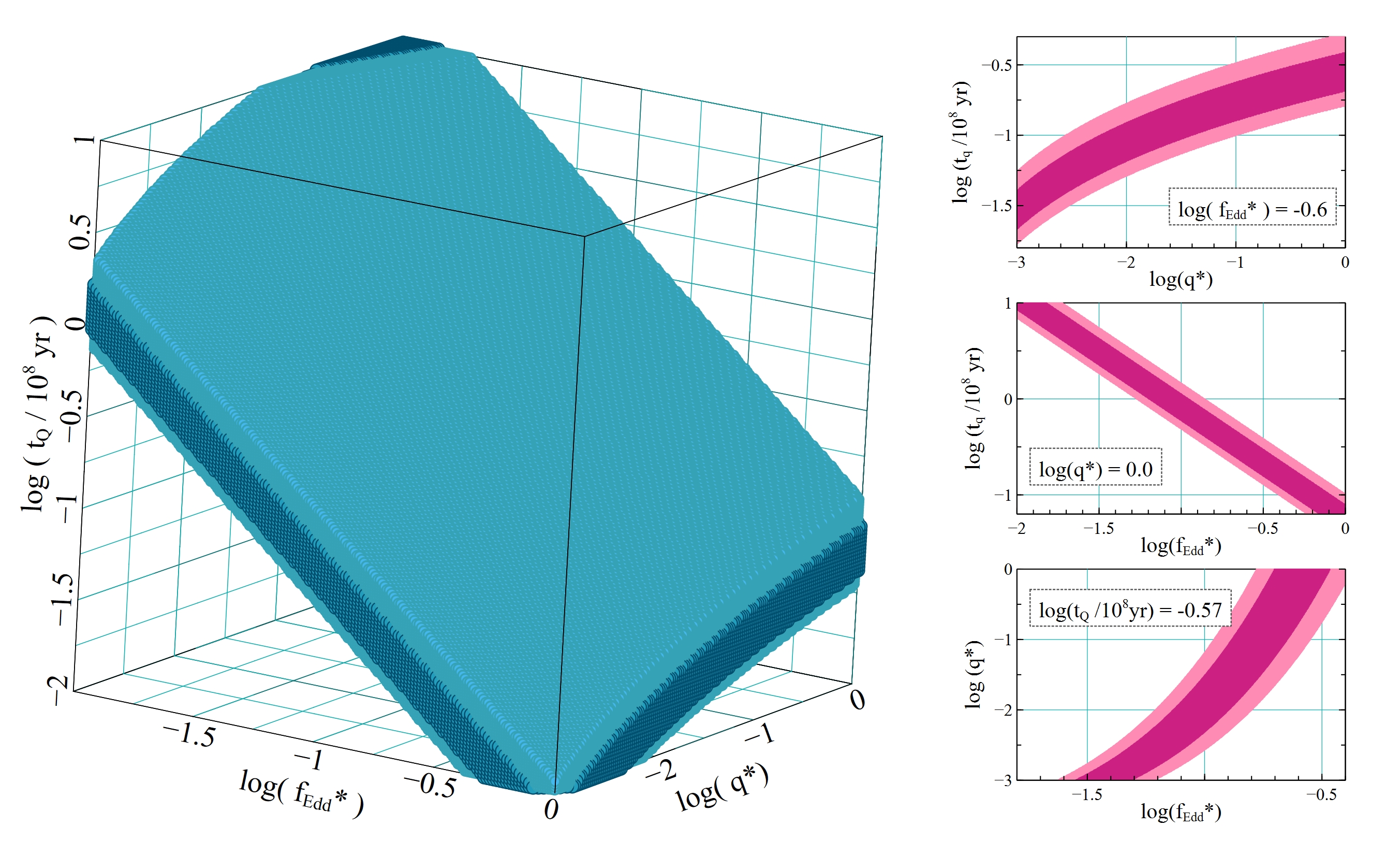}
\caption{Left panel: $68 \%$ and $95 \%$ confidence levels of the posterior in the three-dimensional parameter space of our model, using PTArcade to fit NANOGrav's 15-yr GWB data. Right panels: Two dimensional cross-sections of the posterior with the same confidence levels, and with the third parameter fixed at its fiducial value, $\textrm{log}({f_{\rm Edd}}^*) = -0.6$, $\textrm{log}(q) = 0.0$, and $\textrm{log}(t_{\rm Q}\, /10^8 \, \textrm{yr} )= -0.57$.  There is a trivial log-linear degeneracy between the Eddington ratio $f_{\rm Edd}^*$ and the typical quasar lifetime $t_{\rm Q}$, while the degeneracies involving the mass ratio $q$ are more complicated.
}
\label{fig:combination}
\end{figure}

The left panel of Figure~\ref{fig:combination} shows the allowed combinations in our three dimensional parameter space, utilizing the likelihoods obtained with the public Monte-Carlo Markov Chain (MCMC) package PTArcade ~\citep{mitridate_2023,Mitridate:2023oar}.
The light-coloured region marks the $68 \%$, and the dark coloured region marks the $95 \%$ confidence-level fit to the NANOGrav data.   
In the right panels of Figure~\ref{fig:combination}, we show the two-dimensional cross sections with the same confidence levels, with the third parameter fixed at its fiducial value (i.e. $\textrm{log}({f_{\rm Edd}}^*) = -0.6$, $\textrm{log}(q) = 0.0$, and $\textrm{log}(t_{\rm Q}\, /10^8 \, \textrm{yr} )= -0.57$).

There is a trivial log-linear degeneracy between the Eddington ratio $f_{\rm Edd}^*$, and the typical quasar lifetime $t_{\rm Q}$, since the GWB amplitude scales as a power-law with these parameters. The degeneracies involving the mass-ratio $q$ are more complicated, due to the non-linear dependence of the chirp-mass on $q$.    Our modeling is very simple, and can be expanded in the future to include different shapes for the assumed distributions, and allowing their dispersions to be free parameters, as well as incorporating dependencies on redshift and SMBH mass.  Nevertheless, this figure shows that there are strong degeneracies between parameters, which make many combinations viable fits to the data. 

\subsection{Contribution of SMBHBs with different masses and redshifts}

\begin{figure}[t]
\includegraphics[width=\textwidth]{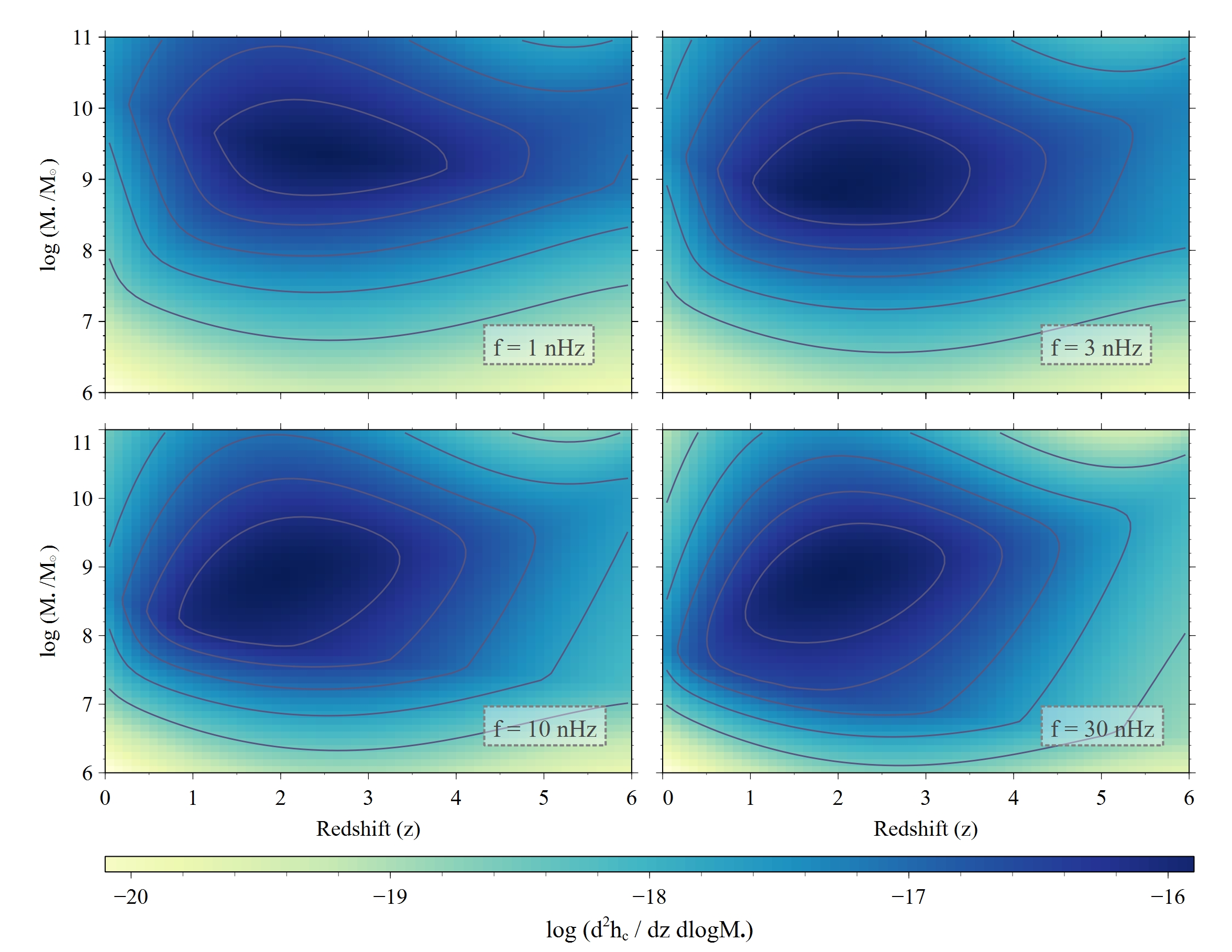}
\caption{Contributions to the present-day GWB of SMBHBs with different masses and redshifts at the GW frequencies of $f=1$, 3, 10 and 30 nHz. The bin-size of the two-dimensional distributions in each panel is: $\Delta z \times \Delta \log(M_{\bullet} / M_{\odot})=0.1 \times 0.1$. 
Binaries in the $M_{\bullet} \sim 10^8 - 10^{10} \, M_{\odot}$ and z $\sim 1- 3 $ ranges dominate, with the contributions shifting toward higher masses and slightly higher redshifts at lower frequencies. These trends are driven in most part by the shape and redshift-evolution of the quasar LF.} 
\label{fig:contribution}
\end{figure}

The present-day gravitational wave background is an accumulation of the GWs from mergers over cosmic history, by SMBH binaries with all masses. Figure \ref{fig:contribution} shows the contributions of binaries in different mass and redshift ranges at four different GW frequencies in our fiducial model. In each panel, it is clear that the most significant contributions are from high-mass SMBHs with masses of $10^8\lsim M_{\bullet}/{\rm M_{\odot}} \lsim  10^{10}$ in the redshift range $1\lsim z \lsim 3 $. This result is unsurprising and follows directly from the evolving QLF, which peaks in the same redshift range~\citep{Richards+2006}. Additionally, the most massive binaries produce the strongest GWs, but the abundance of SMBHs decreases steeply for masses above $\gsim  10^{10} \, {\rm M_{\odot}}$. 

Taking a closer look at the four different panels, small disparities can be seen, with contributions shifting toward higher masses and higher redshifts at lower frequencies. In the $f=1$ nHz bin, the most significant contribution are from binaries with even higher masses of 
$10^9\lsim M_{\bullet}/{\rm M_{\odot}} \lsim  10^{10}$ at slightly higher redshifts  $1.3 \lsim z \lsim 3.8$. By comparison, binaries
with $10^8\lsim M_{\bullet}/{\rm M_{\odot}} \lsim  10^{9}$ and   $1\lsim z \lsim 3$ dominate at $f=30$ nHz. Again, this is unsurprising, given the inverse relation between the characteristic GW frequency and chirp mass, and the shape of the evolving QLF (with the abundance of more luminous quasars peaking at higher redshifts).   However, this result differs from other models in the literature, in which the contribution to the present-day GWB is concentrated at lower redshifts (see, e.g. Ref.~\citealt{NANOGrav15-SMBHBs} and references therein).

\subsection{The effect of discreteness in the SMBHB population and the GWB spectrum}

Our model assumes that the supermassive black hole binary (SMBHB) population is continuously sampled across the parameter space, with the number of binaries contributing in each (mass, redshift and frequency) bin equal to the mean expected value. While this is a good approximation at lower frequencies, where each bin contains $\gg$$1$ sources on average, Ref.~\citep{Sesana:2008} demonstrated that at higher frequencies ($f \gsim 1~{\rm yr^{-1}}$), the number of binaries contributing significantly to the GWB in a given frequency bin can decrease and approach unity (for their population model). Any single realization of the SMBHB population will then produce different discrete numbers of binaries in these bins, with large variations from one realization to another.  Ignoring this discreteness leads to missing fluctuations across frequency in the GWB spectrum, and typically (in most random realizations) an overestimation of the resulting GWB amplitude  (see also refs.~\citealt{KocsisSesana2011} and \citealt{DvorkinBarausse2017} for demonstrations of these points). The frequency at which these effects become significant, and the level of its impact on the GWB amplitude and spectrum, depend on the number and distribution of both lower- and higher-mass SMBHBs in different models. We analyzed the number of sources within specific frequency, total mass, and redshift bins in our model to evaluate the extent to which discreteness influences our results.
Specifically, we follow ref.~\citep{Sesana:2008} and compute $\tilde{M}$ in each $(z,f)$ bin such that the number of sources in that bin with mass greater than $\tilde{M}$ equals $1$. We then re-compute $h_c(f)$ assuming that sources above $\tilde{M}$  are not present in the data (as would be the case for $\approx 40\%$ of the realizations, assuming a Poisson distribution in each bin).

\begin{figure}[t]
\includegraphics[width=\textwidth]{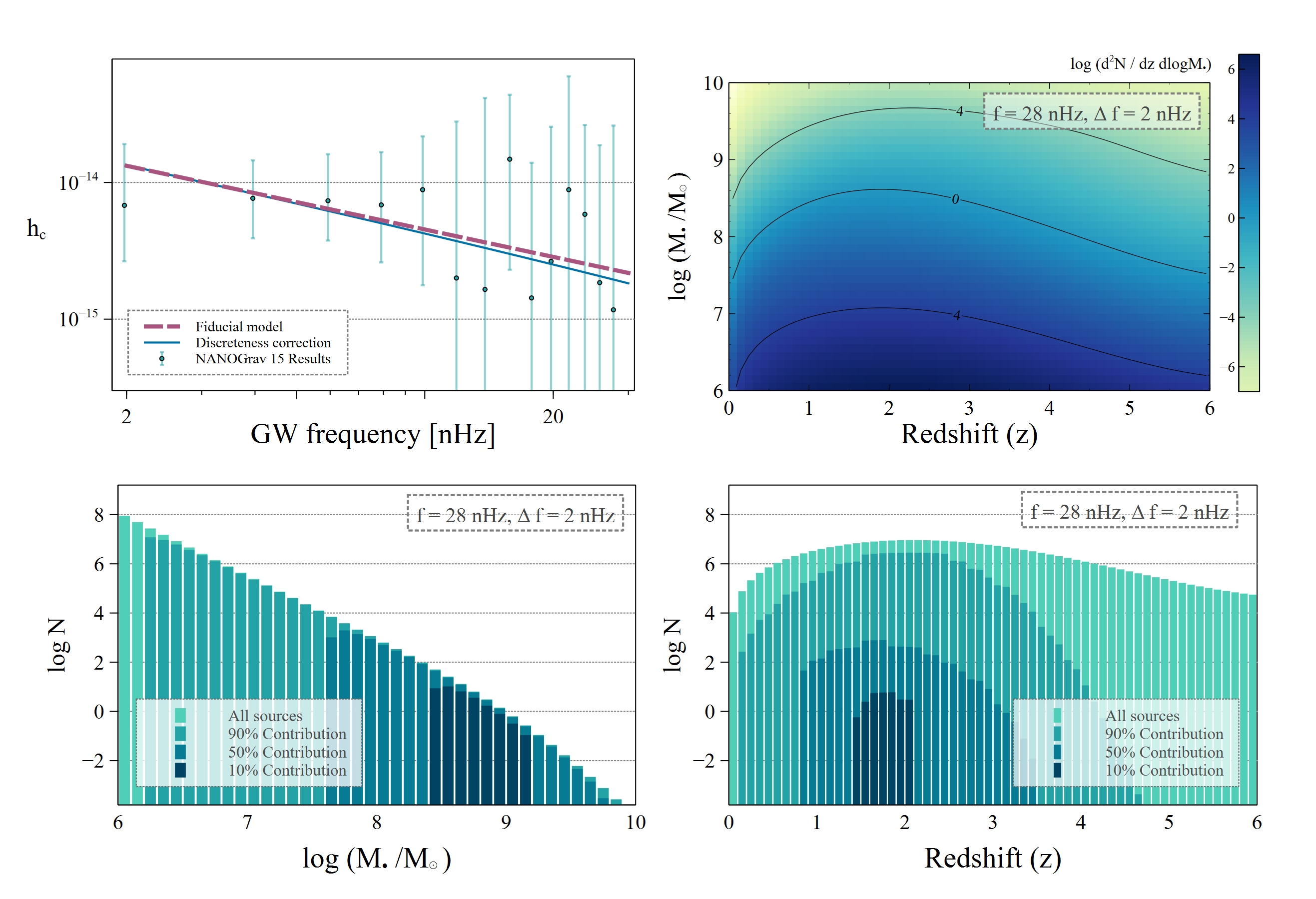}
\caption{Top left: The GWB amplitude in the fiducial model compared to a model corrected for discreteness effects. In the high-frequency regime, the spectrum typically steepens due to the low number of sources contributing in higher-frequency bins. However, this steepening is not as pronounced as in previous work, because the GWB is composed of a larger number of more distant sources in our case.
Top right: A two-dimensional map showing the number of SMBHBs contributing to the GWB at $f = 28 \pm 1$ nHz across different total mass and redshift bins.
Bottom left: The distribution of SMBHB sources as a function of the total mass of the binaries. In addition to the total sources, this panel also marks the top $10\%$, $50\%$ and $90\%$ contributing sources in different colors, as labeled.
Bottom right: The distribution of SMBHB sources across different redshift intervals. Similarly to the previous panel, it presents the top $10\%$, $50\%$ and $90\%$ contributors to the GWB signal. 
}
\label{fig:discreteness}
\end{figure}

The top left panel of Fig.~\ref{fig:discreteness} illustrates that incorporating the discreteness effect leads to a slight deviation from the originally predicted power-law GWB spectrum. A steepening of the spectrum is apparent at higher frequencies ($f \geq 10$ nHz), though the effect is relatively minor compared to the earlier studies mentioned above. Additionally, the steepen spectrum remains well within the errors of the NANOGrav result.
The remaining panels of Fig.~\ref{fig:discreteness} present our calculations of source numbers in various total mass and redshift bins. Although computations were performed for multiple frequency bins, we focus here on the central frequency $f = 28$ nHz, which is closest to the highest accessible frequency in the NANOGrav results. At this frequency, the impact of discreteness corrections is most pronounced. The frequency bin width is consistent with the 15-year observational baseline, $\Delta f = 1/T \simeq 2$ nHz. 

The 2D map in the upper right panel of Fig.~\ref{fig:discreteness} shows the distribution of the number of SMBHBs in our model within the range $10^6 M_{\odot} \leq M_{\bullet} \leq 10^{10} M_{\odot}$ and $0 \leq z \leq 6$.  The bin widths are $\Delta M_{\bullet} = 0.1 , \log(M/M_{\odot})$ and $\Delta z = 0.1$, respectively. 
The bottom panels project the distribution of sources by total black hole mass (left panel) and redshift (right panel). Thesese panels also depict the distribution of sources that contribute most significantly to the GWB. The top $10\%$, $50\%$ and $90\% $ contributors are highlighted in different colors.
The results indicate that the continuous approximation becomes less valid only in the highest mass regime ($M_{\bullet} \geq 10^9 {\rm M_{\odot}}$). However, this produces only a modest deviation of the overall GWB spectrum, as there are still tens of sources around $10^9 {\rm M_{\odot}}$ and $z\approx 2$ which dominate this frequency bin (see Fig.~\ref{fig:contribution}), and 
lower-mass binaries contribute to the GWB through in even larger numbers.

We conclude that our model differs from other models in the literature, in that the GWB is composed of a larger number of sources, associated with quasars at higher redshifts.   As a result, the spectral slope is not as steepened at high frequencies as in previous models, and also should remain smoother. 
This conclusion is perhaps not surprising: the QLF has a much slower power-law decrease at the bright end than the exponential decrease at the high-mass end of spheroid mass- or velocity-functions.  Indeed, in a differential version of the 'Soltan argument' \cite{Soltan1982} that compares quasar light and remnant SMBH masses as a function of quasar luminosity, this necessitates average radiative efficiencies $\epsilon=L_{\rm bol}/\dot{M}_{\rm bh}c^2$ for bright quasars that are a factor of several above the canonical value of $\epsilon=10\%$~(\cite{YuTremaine+2002}; see also \cite{Haiman+2007}).
The relatively large number of bright quasars, peaking at $z=2-3$, makes it less surprising that the SMBHS associated with this quasars dominate the GWB in our model, compared to fewer and lower-redshift sources in previous models.
A possible test of our model is that we expect to find typically a lower lever of angular anisotropy, compared to previous models in which the GWB is composed of fewer and more nearby sources.

\section{Discussion}
\label{sec:discussion}

Our main results in this study is that associating SMBH binaries with quasars, with very simple assumptions, yields good fits to the observed GWB.
Quasars have been connected to modeling the GWB in the PTA bands in the past, both directly as well as indirectly.  
As an example of the latter, ref.~\citet{Izquierdo-Villalba+2022} used a semi-analytical model to populate the dark matter halo merger trees in the Millennium simulations, to model the PTA background. In their model, they also apply a detailed prescription for BH accretion and resulting AGN activity, and predicted the quasar luminosity function (QLF) between $0\lsim z \lsim 3$.  They find a reasonable fit to the QLF, although typically their model overpredicts the observed abundance of the most luminous quasars, whose SMBHs are responsible for the GWB (see their Fig.11).
In a similar vein, ref.~\citet{Barausse+2023} have used semi-analytic models to follow the formation and evolution of massive black hole binaries. They showed that when their models are calibrated against the recently measured PTA GWB, these models are in reasonable agreement with the measured QLF at $z\sim 0$ (see their Fig.3).

A more direct prediction of the GWB, starting from quasars, was made in ref.~\cite{Casey-Clyde+2022}, in an approach similar to ref.~\cite{Haiman+2009} and the one we also followed here. In their study, 
the starting point is the quasar activation rate from refs.~\cite{Hopkins+2006,Hopkins+2007}, which is based on the convolution of an empirical QLF with parameterized quasar light-curves calibrated to hydrodynamical simulations. This is then converted to the SMBHB merger rate and used to predict the GWB. By comparison, we adopted a recently updated parameteric QLF, and replaced the above by directly specifying $f_{\rm Edd}$ (or its distribution) and $t_{\rm Q}$ as free model parameters.   Although not anchored to simulations, our simpler parameterization makes the  values of these basic parameters preferred by the NANOGrav data more transparent. Ref.~\cite{Casey-Clyde+2022} fit their model to the GWB amplitude $A_{\rm 1yr}=(1.9\pm 0.4)\times 10^{-15}$, based on the earlier NANOGrav 12.5-year data release~\cite{NANOGrav12.5}, which is only $\sim20$\% lower than the more recent best-fit value of $A_{\rm 1yr}=2.4\times 10^{-15}$ we used in this paper (and well within the errors).   In terms of our results, ref.~\cite{Casey-Clyde+2022} find that the abundance of high- mass ($M_{\bullet}/M_{\odot} > 10^{8.5} $) SMBHBs is roughly four times higher than the number of quasars in the local universe, and that the largest contribution to the GWB is from $z\sim 0.5$.  In contrast, in our fiducial model we set $f_{\rm bin}=1$ -- i.e. we assume a simple one-to-one correspondance between quasars and SMBHBs, so they have the same abundance. 
We found that the GWB amplitude can be fit with reasonable combinations of the other parameters, particularly $t_{\rm Q}$ -- which is trivially degenerate with $f_{\rm bin}$, i.e. we constrain only their ratio $t_{\rm Q}/f_{\rm bin}$. 
We find that the difference between the two approaches can not be attributed to the different adopted QLFs. The QLF based on \cite{Hopkins+2007} and used in \cite{Casey-Clyde+2022} predicts somewhat more quasars at the high-luminosity end (and fewer lower luminosity sources) than the QLF based on \cite{Kulkarni+2019} we used.  We have verified that if we repeat our analysis, but switch to the QLF from ref.~\cite{Hopkins+2007}, this produces a  higher GWB amplitude.  In other words, switching to this QLF does not necessitate more mergers than available from quasars, to explain the GWB.  We conclude that the different choices of the distributions of $f_{\rm{Edd}}$, $q$ and especially in the quasar lifetime $t_{\rm Q}$ must be responsible for the discrepancy in the final outcomes.  

Overall, our results suggest a consistency between the GWB amplitude and shape, and the assumption that a significant fraction of quasars are associated with SMBHB's, with the contribution to the GWB peaking at $z\sim 2$, near the peak of the QLF.  This conclusion further depends, however, on our assumption (also made in the quasar-based model in ref.~\cite{Casey-Clyde+2022}) that SMBHB coalescences and bright quasar phases are contemporaneous.  Whether this is the case is not clear -- the relatively brief bright quasar phases and the similarly short duration of the SMBHBs in the PTA band may be significantly separated in time (with either one preceding the other).

In a recent follow-up study, ref.~\cite{Casey-Clyde+2024} have further assessed whether SMBHBs are preferentially found in quasars compared to random galaxies, using the recent GWB measurement by NANOGrav as one of their model constraints.  They concluded that quasars are at most $\sim$ an order of magnitude more likely than a random galaxy to host an SMBHB.  This naively implies that in our model the binary fraction is at most $f_{\rm bin}\sim0.1$, which would then have to be absorbed by a corresponding order-of-magnitude adjustment of the best-fit combination of the remaining parameters $f_{\rm Edd}$, $t_{\rm Q}$ and $q$.

We also note that the SMBH binaries that produce the GWB have orbital periods of order a year.  If these binaries are luminous quasars, they are expected to exhibit periodic variability on timescales of order the orbital time either from hydrodynamical effects (see, e.g. \citealt{Westernacher+2022} and references therein) from relativistic Doppler modulation (see, e.g. \citealt{D'Orazio+2015}), or from self-lensing~\citep{Haiman2017,D'OrazioDiStefano2018,Pihajoki+2018,Ingram2021,DavelaarHaiman2022}.   Searches for such variability have yielded a sample of several dozen SMBH binary candidates in large existing time-domain surveys, including 
the Catalina Real-time Transient Survey (CRTS; \citealt{Graham+2015}), 
the Palomar Transient Factory (PTF; \citealt{Charisi+2016}) and 
the Zwicky Transient Facility (ZTF; \citealt{Chen+2024}.  In each case, approximately $\sim10^{-3}$ of the full quasar sample was found to exhibit significant periodicity.  This is consistent with the scenario in which most quasars correspond to SMBH mergers, since the GW inspiral time for $\sim10^{9}{\rm M_\odot}$ binaries, from the separation when their orbital period is $\sim$ yr, is $t_{\rm GW}\sim10^5~{\rm yr}$, which is approximately $10^{-3}$ of the quasar lifetime~\citep{Martini2004}.   However, many of these candidates can be false positives due to stochastic brightness variations mimicking periodicities~\citep{Vaughan+2016}, and indeed if all candidates were genuine binaries, the GWB would likely be overproduced~\citep{Sesana+2018}.   Overall, because the time-domain quasar samples suffer from several selection effects, it is currently difficult to use these results to establish a reliable connection between quasars and SMBH binaries. However, this situation should improve significantly in large forthcoming time-domain surveys, such as the Vera Rubin Observatory's LSST \citep{LSST2021}, which should be able to unambiguously identify SMBH binaries.

Our modeling in this work is simplified, with many possible future improvements.  One particular caveat is that there could be many highly dust-obscured quasars which are not accounted for in the quasar LF we adopted; in fact the majority of the most luminous quasars may be obscured (see, e.g. Ref.~\citealt{Assef+2015} and references therein).  This would increase the predicted GWB amplitude, or, alternatively, decrease the required $f_{\rm bin}$ and/or yield a correspondingly adjusted best-fit combination for $f_{\rm Edd}$, $t_{\rm Q}$ and $q$.

\section{Conclusions}
\label{sec:conclusions}

In this paper, we hypothesized that there is a one-to-one correspondance between luminous quasars and coalescing SMBH binaries.  Using a simple model to codify this connection, we used the empirically measured quasar luminosity function (QLF) to predict the present-day stochastic gravitational wave background (GWB).
This approach bypasses the need to model the cosmological evolution of SMBH or galaxy mergers from simulations or semi-analytical models. 
Although our modeling is simplified, it demonstrates that a scenario in which a significant fraction of bright quasars are activated by galaxy mergers, which also produce promptly coalescing SMBH binaries, is consistent with the measured GWB.  The GWB in this case is dominated by a relatively large number of distant $\sim10^9{\rm M_\odot}$ SMBHs at $z\approx 2-3$, at the peak of quasar activity.  This differs from most other models in the literature, in which the GWB is dominated by fewer SMBH coalescences at lower redshifts, and would have implications for several aspects of the GWB, such as its stochasticity and angular anisotropy (both of which is less pronounced in our case), and the ability of PTAs to resolve individual SMBHBs (which will be more difficult for the more distant sources in our case).

\section*{Acknowledgments}

We thank Chengcheng Xin and Girish Kulkarni for useful discussions.
ZH gratefully acknowledges the hospitality of E\"otv\"os University during an extended sabbatical visit, where this work began.
ZH acknowledges support from NSF grant AST-2006176 and NASA grants 80NSSC22K0822 and 80NSSC24K0440.

\vspace{2\baselineskip}
\bibliography{refs}

\end{document}